\documentclass[twocolumn,prb]{revtex4}
\usepackage{amsfonts}
\usepackage[T1]{fontenc}
\usepackage{amsmath,amsbsy,amssymb,graphicx}
\usepackage{times}

\begin{document}

\title{ Magnetic second-order topological insulators and semimetals}
\author{Motohiko Ezawa}
\affiliation{Department of Applied Physics, University of Tokyo, Hongo 7-3-1, 113-8656,
Japan}

\begin{abstract}
We propose magnetic second-order topological insulators (SOTIs).
First, we study a three-dimensional model. It is pointed
out that the previously proposed topological hinge insulator has actually
surface states along the [001] direction in addition to hinge states. We gap
out these surface states by introducing magnetization, obtaining a SOTI only
with hinge states. 
The bulk topological number is the $Z_2$ index protected by the combined symmetry of the four-fold rotation and the inversion symmetry. 
We next study two dimensional magnetic SOTIs, where the
corner states are robust also in the presence of the magnetization. Finally,
we construct a magnetic second-order topological semimetals by layering the
two-dimensional magnetic SOTIs, where hinge-arc states are robust also in
the presence of the magnetization.
\end{abstract}

\maketitle

\textit{Introduction:} Topological insulators (TIs) have opened a new world
in condensed matter physics\cite{Hasan,Qi}. The first example is the quantum
Hall insulator, where the topological number is the Chern number\cite{TKNN}.
It is not necessary to have any symmetries for the quantization of the Chern
number. The current movement of the topological insulator has started with
the time-reversal invariant topological insulator\cite{KaneMele,FuKane},
where the topological number is the $Z_{2}$ index protected by the time
reversal symmetry (TRS). Then, it is generalized to the topological
crystalline insulator, where the mirror Chern number is the topological
number\cite{TCI,Morimoto,Shiozaki}. Here the mirror symmetry protects the
topological phase. Recent interest is renewed in topological insulators
protected by more general crystalline symmetries including the rotational
symmetry\cite{Po2,Po3,ZSong,FuRot,CFang,k4}.

Higher-order topological insulators are extension of the TIs\cite%
{Fan,Science,APS,Peng,Lang,Song,Bena,Schin,Liu,EzawaKagome,EzawaPhos}, to
which the conventional bulk-boundary correspondence is generalized. Here we
focus on three dimensional (3D) crystals. Then, the second-order TI (SOTI)
has 1D topological boundary states (hinge states) but has no 2D topological
boundary states (surface states), while the third-order TI has 0D
topological boundary states (corner states) but has neither surface states
nor hinge states. There are several works to classify them based on
crystalline symmetries\cite{Lang,k4,Gei,Kha}. As a closely related concept
to the SOTI, there are topological hinge insulators\cite{Schin}. They are
TIs possessing topological hinge states. Note that they may have topological
surface states in addition to hinge states. An example\cite{Schin} was
constructed by adding a nontrivial mass term to a 3D TI and by gapping out
some topological surface states. Indeed, when we consider a cube parallel to
the $x$, $y$ and $z$ axes in this example, there appear two surface states
perpendicular to the $z$ axis in addition to four hinge states: See Fig.\ref%
{FigCube}(b).

\begin{figure}[t]
\centerline{\includegraphics[width=0.5\textwidth]{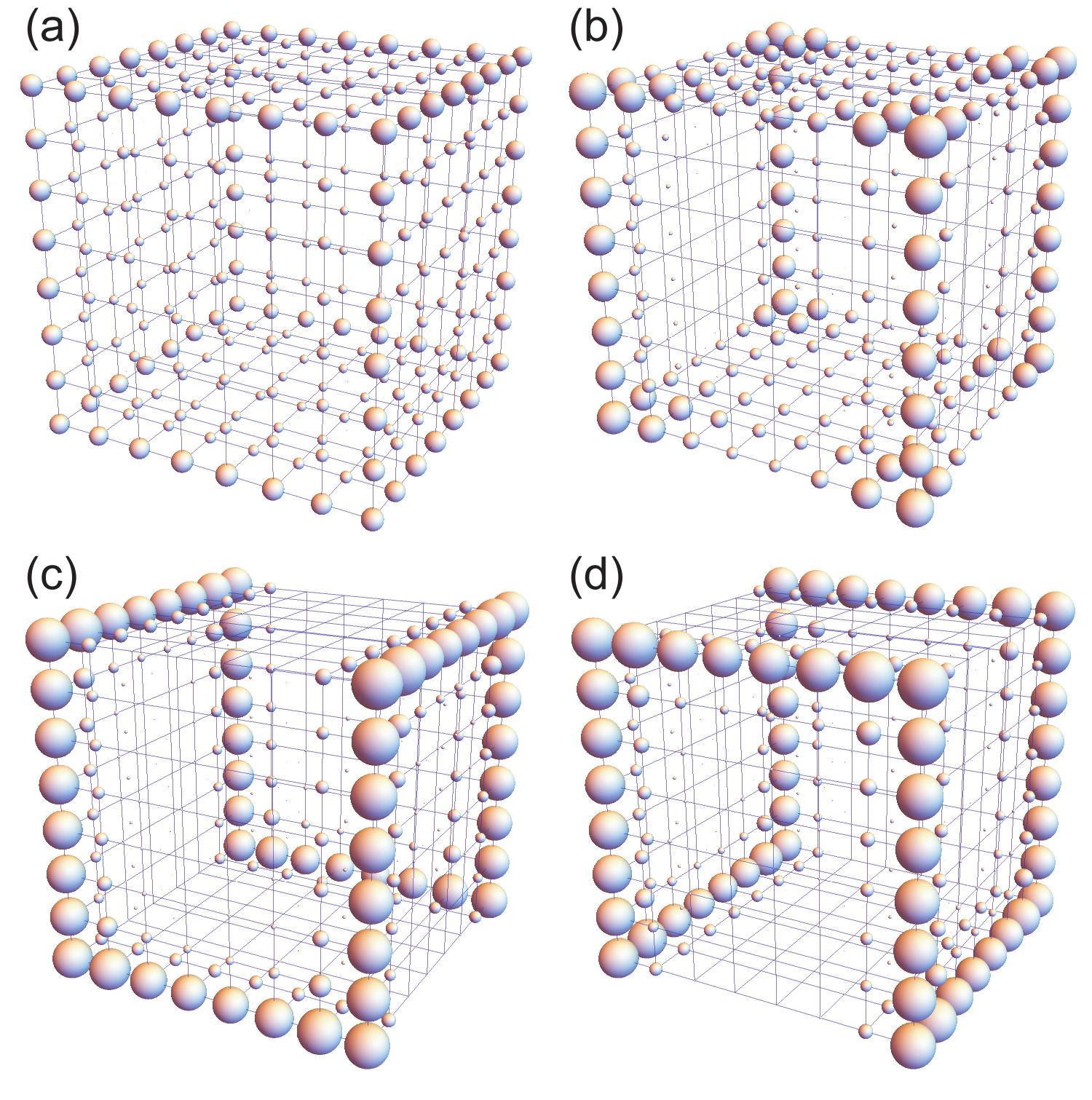}}
\caption{3D SOTI. The real-space plot of the square root of the local
density of states $\protect\sqrt{\protect\rho _{i}}$ for a cube in the case
of (a) the TI, (b) a topological hinge insulator, (c) and (d) magnetic SOTIs
in the presence of magnetization with $B>0$ and $B<0$, respectively. The
amplitude is represented by the radius of spheres. The size of the cube is $%
L=8$. The local density of states becomes arbitrarily small for $L\gg 1$
except for (a) 6 surfaces, (b) 2 surfaces and 4 hinges, (c) and (d) for
hinges.}
\label{FigCube}
\end{figure}

In this paper, introducing magnetization along the $z$ axis, first we
propose a 3D magnetic SOTI by gapping out all the surface states in the
topological hinge insulator\cite{Schin} just mentioned above. As we see in
Fig.\ref{FigCube}(c) and (d), there appear only hinge states without surface
states in the presence of the magnetization. The bulk topological number is
shown to be the $Z_{2}$ index protected by the rotoinversion symmetry $\bar{C%
}_{4}=C_{4}I$, which is the combined symmetry of the four-fold rotation $%
C_{4}$ and the inversion $I$. Second, we construct a 2D magnetic SOTI, where
topological corner states appear. Finally, we construct a magnetic
second-order topological semimetals based on the stacking of the 2D magnetic
SOTI, where hinge-arc states emerge connecting the gap closing points.

\textbf{3D magnetic SOTI:} The typical model for the 3D TI is given by\cite%
{BHZ} 
\begin{equation}
H_{0}=\left( m+t\sum_{i}\cos k_{i}\right) \tau _{z}\sigma _{0}+\lambda
\sum_{i}\sin k_{i}\tau _{x}\sigma _{i}  \label{H0}
\end{equation}%
on the cubic lattice, where $i$ runs over $x,y,z$. It describes\cite{Legner}
topological Kondo insulators SmB$_{6}$. The $\sigma _{i}$ represent the
Pauli matrices corresponding to the spin degrees of freedom and $\sigma _{0}$
is the two by two identity matrix, while $\tau _{i}$ are the Pauli matrices
corresponding to the orbital degrees of freedom. It has the TRS. Protected
by the TRS, it has topological surface states in accord with the
bulk-boundary correspondence as in Fig.\ref{FigCube}(a).

To gap out them by breaking the TRS, the following extra term has been
proposed\cite{Schin}, 
\begin{equation}
H_{\Delta }=\Delta \left( \cos k_{x}-\cos k_{y}\right) \tau _{y}\sigma _{0}.
\label{H1}
\end{equation}%
Additionally we introduce the Zeeman term induced by magnetization,%
\begin{equation}
H_{Z}=B\tau _{0}\sigma _{z}.  \label{HZ}
\end{equation}%
We study the effect of the Zeeman term in the 3D topological hinge
insulators in the following order.

\begin{figure}[t]
\centerline{\includegraphics[width=0.5\textwidth]{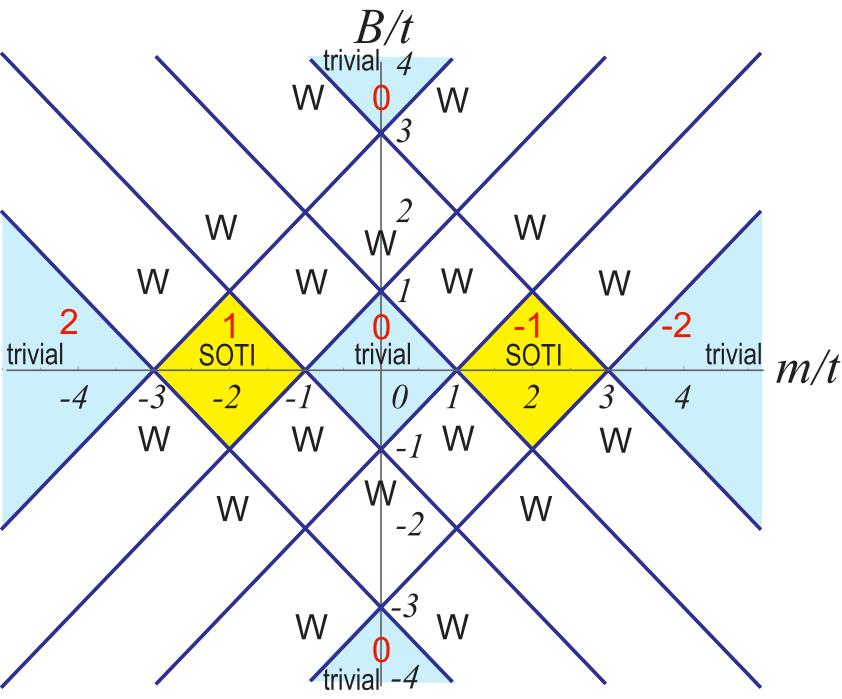}}
\caption{3D SOTI. Topological phase diagram in the $(m/t,B/t)$ plane.
Numbers in red represent the $Z$ index $\protect\kappa _{4}$, which gives
the topological number $\protect\nu $ by the formula $\protect\nu =\text{mod}%
_{2}\protect\kappa _{4}$. The symbol $W$ stands for Weyl semimetal phases.
The SOTI phases are masked in yellow.}
\label{FigPhase}
\end{figure}

\textit{Topological phase diagram:} The band structure is obtained by
diagonalizing the Hamiltonian $H=H_{0}+H_{\Delta }+H_{Z}$. It follows that
the phase boundaries are given by solving the zero-energy condition ($E=0$)
at the four high-symmetry points $\Gamma =\left( 0,0,0\right) $, $S=\left(
\pi ,\pi ,0\right) $, $Z=\left( 0,0,\pi \right) $ and $R=\left( \pi ,\pi
,\pi \right) $ with respect to the four-fold rotation. The energies at these
points are analytically given by%
\begin{eqnarray}
E\left( 0,0,0\right) &=&3t+m\pm B,\quad -3t-m\pm B, \\
E\left( \pi ,\pi ,0\right) &=&t-m\pm B,\quad -t+m\pm B, \\
E\left( 0,0,\pi \right) &=&t+m\pm B,\quad -t-m\pm B, \\
E\left( \pi ,\pi ,\pi \right) &=&3t-m\pm B,\quad -3t+m\pm B.
\end{eqnarray}%
We show the phase diagram in Fig.\ref{FigPhase}. In the absence of the
Zeeman term\cite{Schin}, the system is topological for $1<\left\vert
m/t\right\vert <3$ and trivial for $\left\vert m/t\right\vert <1$ or $%
\left\vert m/t\right\vert >3$: See Fig.\ref{FigPhase}. Insulators emerge in
the region including the phases with $B=0$. Later we identify the bulk
topological number $\nu $ and find that it changes its value at these phase
boundaries: See (\ref{TC-C4I}). The phase diagram consists of topological
and trivial insulator phases and Weyl semimetal phases.

\begin{figure}[t]
\centerline{\includegraphics[width=0.5\textwidth]{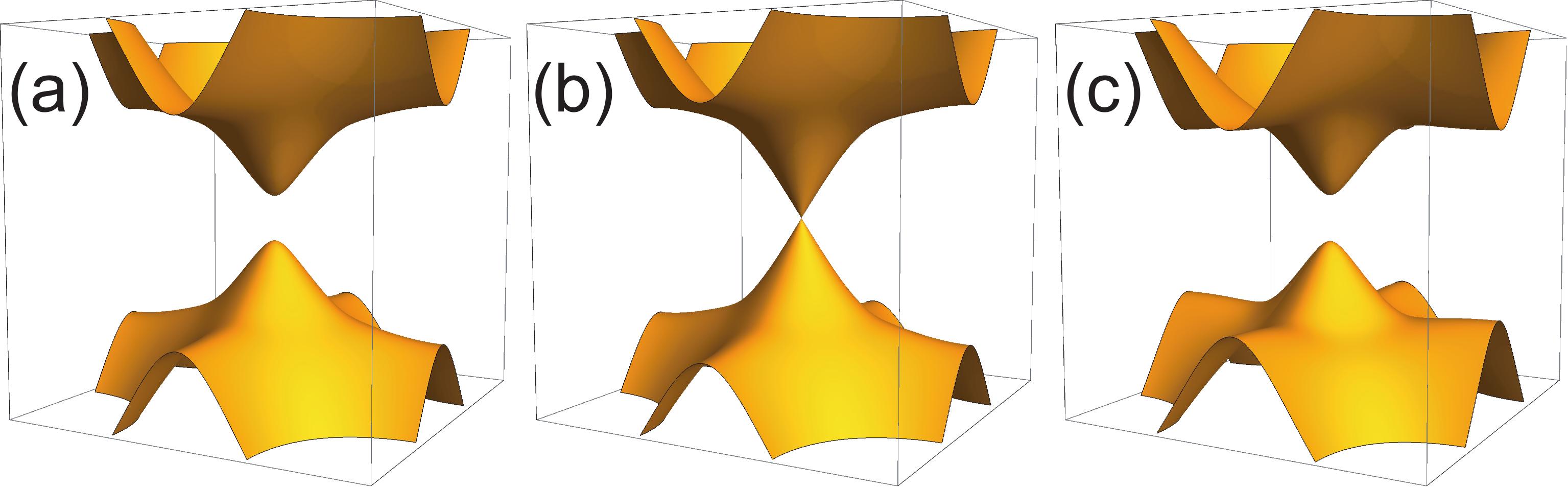}}
\caption{3D SOTI. Surface band structure of a thin film. Surface states of
the topological hinge insulator. (a) along the [100] and [010] direction;
(b) those along the [001] direction, where the gap closes at the $M$ point
in the surface Brillouin zone; (c) Surface states of the magnetic SOTI along
the [001] direction with $B=t/2$. The surface states are gapped in the
presence of magnetization.}
\label{FigSurface}
\end{figure}

\textit{Symmetries:} In order to identify the bulk topological invariant, it
is necessary to study the symmetry of the Hamiltonian $H_{0}$. We note that $%
TH_{0}\left( \mathbf{k}\right) T^{-1}=H_{0}\left( -\mathbf{k}\right) $ and $%
IH_{0}\left( \mathbf{k}\right) I^{-1}=H_{0}\left( -\mathbf{k}\right) $,
where $T=\tau _{0}\sigma _{y}K$ generates the TRS with the complex
conjugation $K$, while $I=\tau _{z}\sigma _{0}$ the inversion symmetry. In
addition, there is a four-fold rotational symmetry $C_{4}$, 
\begin{equation}
C_{4}H_{0}\left( k_{x},k_{y},k_{z}\right) C_{4}^{-1}=H_{0}\left(
-k_{y},k_{x},k_{z}\right) ,
\end{equation}%
where 
\begin{equation}
C_{4}=\tau _{0}\exp \left[ -\frac{i\pi }{4}\sigma _{z}\right] .
\end{equation}%
It is the generator of the $\pi /4$ rotation. The $H_{\Delta }$ breaks both
of the TRS and the inversion symmetry but preserves\cite{Schin} the combined
symmetry $C_{4}T$ and the rotoinversion symmetry $\bar{C}_{4}=C_{4}I$. Our
concern is the effect of the Zeeman term. The Zeeman term $H_{Z}$ breaks $%
C_{4}T$ but preserves $\bar{C}_{4}$.

\textit{$Z_{2}$ index protected by $\bar{C}_{4}$:} We can define\cite{k4}
the $Z$ index $\kappa _{4}$ protected by the rotoinversion symmetry $\bar{C}%
_{4}$, 
\begin{equation}
\kappa _{4}=\frac{1}{2\sqrt{2}}\sum_{K}\sum_{\alpha }e^{\frac{i\alpha \pi }{4%
}}n_{K}^{\alpha },  \label{Kappa4}
\end{equation}%
where $K$ runs over the high symmetry points $\Gamma $, $S$, $Z$, $R$; $%
n_{K}^{\alpha }$ is the number of the occupied bands with the eigenvalue $e^{%
\frac{i\alpha \pi }{4}}$ of the symmetry operator $\bar{C}_{4}$, $\bar{C}%
_{4}|\psi \rangle =e^{\frac{i\alpha \pi }{4}}|\psi \rangle $. Because of the
relation $\left( \bar{C}_{4}\right) ^{4}=-1$, $\alpha $ is quantized to be $%
\alpha =1,3,5,7$. We explicitly evaluate $\kappa _{4}$ using the formula (%
\ref{Kappa4}), which is shown in Fig.\ref{FigPhase}. It follows that the
topological phases at $B=0$ are extended to the regions with $B\neq 0$, as
shown in Fig.\ref{FigPhase}. When there is the TRS, there is a relation\cite%
{k4} that mod$_{2}\kappa _{4}=\nu _{0}$, where $\nu _{0}$ is the $Z_{2}$
index for the time-reversal invariant topological insulators. Furthermore,
by calculating the band structure of a square prism, we can check that no
hinge states emerge for $B\neq 0$ in the phase indexed by $\kappa _{4}=0,\pm
2$ in the phase diagram (Fig.\ref{FigPhase}). It implies that the bulk
topological index is given by 
\begin{equation}
\nu =\text{mod}_{2}\kappa _{4},  \label{TC-C4I}
\end{equation}%
which is a generalization of $\nu _{0}$ in the absence of the TRS.

\begin{figure}[t]
\centerline{\includegraphics[width=0.5\textwidth]{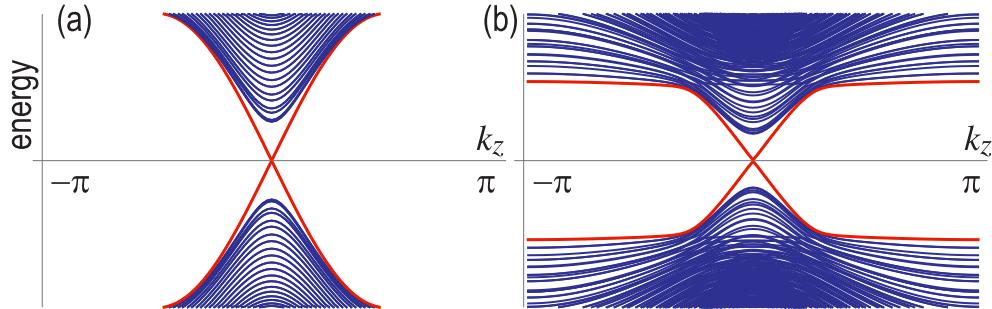}}
\caption{3D SOTI. Band structure of the hinge states (a) without the Zeeman
term and (b) with the Zeeman term $B=t/2$. The hinge states survives even in
the presence of the Zeeman term. The horizontal axis is the momentum $k_{z}$%
. We have set $m/t=2$, $\protect\lambda =t$ and $\Delta =t/4$.}
\label{FigChiralHinge}
\end{figure}

\textit{Surface states:} We study the surface states in the topological
phase. It is shown that the surface states along the [100] and [010]
directions are gapped due to the term $H_{\Delta }$ as in Fig.\ref%
{FigSurface}(a). However, we find the gapless surface states along the [001]
direction as in Fig.\ref{FigSurface}(b). This is because that the $C_{4}T$
and $\bar{C}_{4}$ symmetries are preserved along the [001] direction but
broken along the [100] and [010] directions. Because of the emergence of the
topological surface states, the topological hinge insulator is not a SOTI.
Nevertheless, this gapless surface states can be gapped out by introducing
the Zeeman term as in Fig.\ref{FigSurface}(c). On the other hand, the [100]
and [010] surface states remain to be gapped in the presence of the Zeeman
term.

\textit{Hinge states:} We calculate the band structure of a square prism in
the topological insulator phase to examine the hinge states. The hinge
states remain as they are even in the presence of the magnetization, as
shown in Fig.\ref{FigChiralHinge}. These hinge states are protected by the $%
Z_{2}$\ index associated with the $\bar{C}_{4}$\ symmetry.

\textbf{2D magnetic SOTI:} Next, we study a magnetic SOTI model in two
dimensions. The Hamiltonian is given by setting $i=x,y$ in the Hamiltonian
of the SOTI in three dimensions. The symmetry analysis is almost the same as
in the 3D case just by neglecting the $z$ coordinate. We discuss the
properties of a 2D magnetic SOTI in the following order.

\begin{figure}[t]
\centerline{\includegraphics[width=0.5\textwidth]{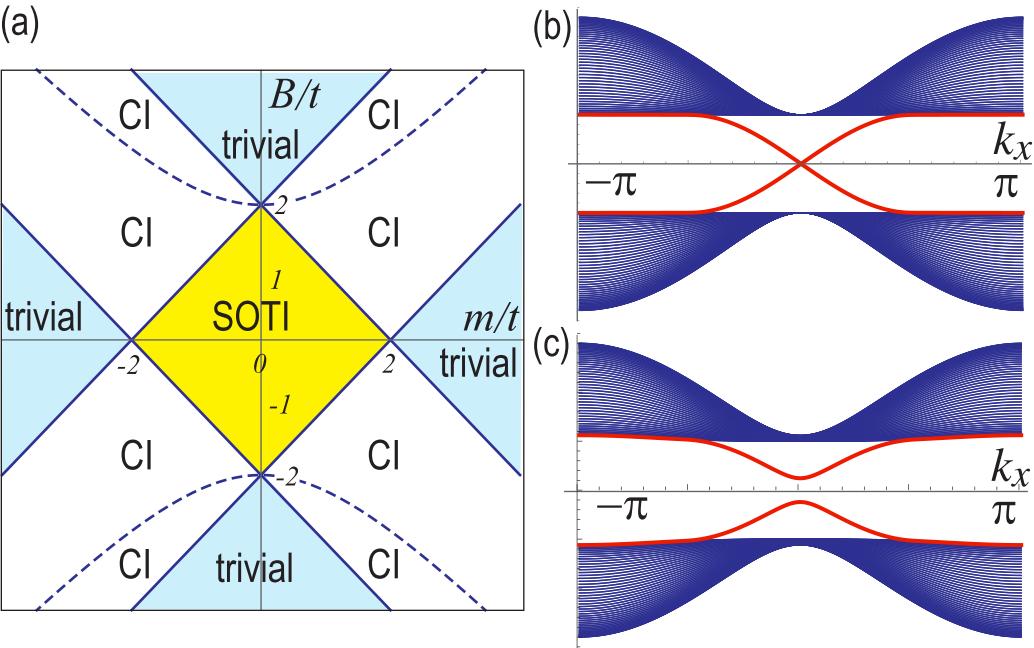}}
\caption{2D SOTI. (a) Topological phase diagram in the $(m/t,B/t)$ plane,
which contains three distinctive phases: the trivial phase, the SOTI phase
and the Chern TI phase (CI). Band structures of (b) a nanoribbon in the
absence of the $H_{\Delta }$ term and (c) the one in the presence of the $%
H_{\Delta }$ term, where we have chosen $\Delta =t/4$. Red curves represent
edge modes in (b). We have set $m= \protect\lambda =t$.}
\label{FigChiralRibbon}
\end{figure}

\textit{Topological phase diagram:} The Brillouin zone is a square with four
corners, $\Gamma =(0,0)$, $X=(\pi ,0)$, $Y=(0,\pi )$ and $M=\left( \pi ,\pi
\right) $. The massive Dirac cone exists at the $M$ point for $|m-2t|<|m+2t|$
and at the $\Gamma $ point for $|m-2t|>|m+2t|$. There are two high-symmetry
points, $\Gamma $ and $M$. At these points the TRS and the $C_{4}$ symmetry
are respected. The energy spectrum reads $E=\pm \left\vert 2t+\eta
m\right\vert $ with the two-fold degeneracy at the $\Gamma $ point with $%
\eta =1$ and at the $M$ point with $\eta =-1$. In the presence of the Zeeman
term, the energies at these points are analytically given by%
\begin{eqnarray}
E\left( 0,0\right) &=&2t+m\pm B,\quad -2t-m\pm B, \\
E\left( \pi ,\pi \right) &=&2t-m\pm B,\quad -2t+m\pm B.
\end{eqnarray}%
The topological phase diagram, given in Fig.\ref{FigChiralRibbon}(a),
consists of a SOTI phase, Chern TI (CI) phases and trivial insulator phases.
We can check there are chiral edge states for nanoribbons in the CI phase.
We have also gap closing in the CI phase at $E\left( \pi ,0\right) =E\left(
0,\pi \right) =0$ with $E\left( \pi ,0\right) =E\left( 0,\pi \right) =\sqrt{%
m^{2}+4\Delta ^{2}}\pm B$, $-\sqrt{m^{2}+4\Delta ^{2}}\pm B$, which are
plotted in dotted curves in Fig.\ref{FigChiralRibbon}(a).

\textit{Edge states:} We calculate the band structure of nanoribbons without
the Zeeman term in the topological phase ($\left\vert m/t\right\vert <2$).
In the absence of the $H_{\Delta }$ term, there are topological edge states
as shown in Fig.\ref{FigChiralRibbon}(b). They are gapped by the $H_{\Delta
} $ term as shown in Fig.\ref{FigChiralRibbon}(c). Namely, edge states are
absent since the rotoinversion symmetry $\bar{C}_{4}$ is broken in the
nanoribbon geometry.

\begin{figure}[t]
\centerline{\includegraphics[width=0.5\textwidth]{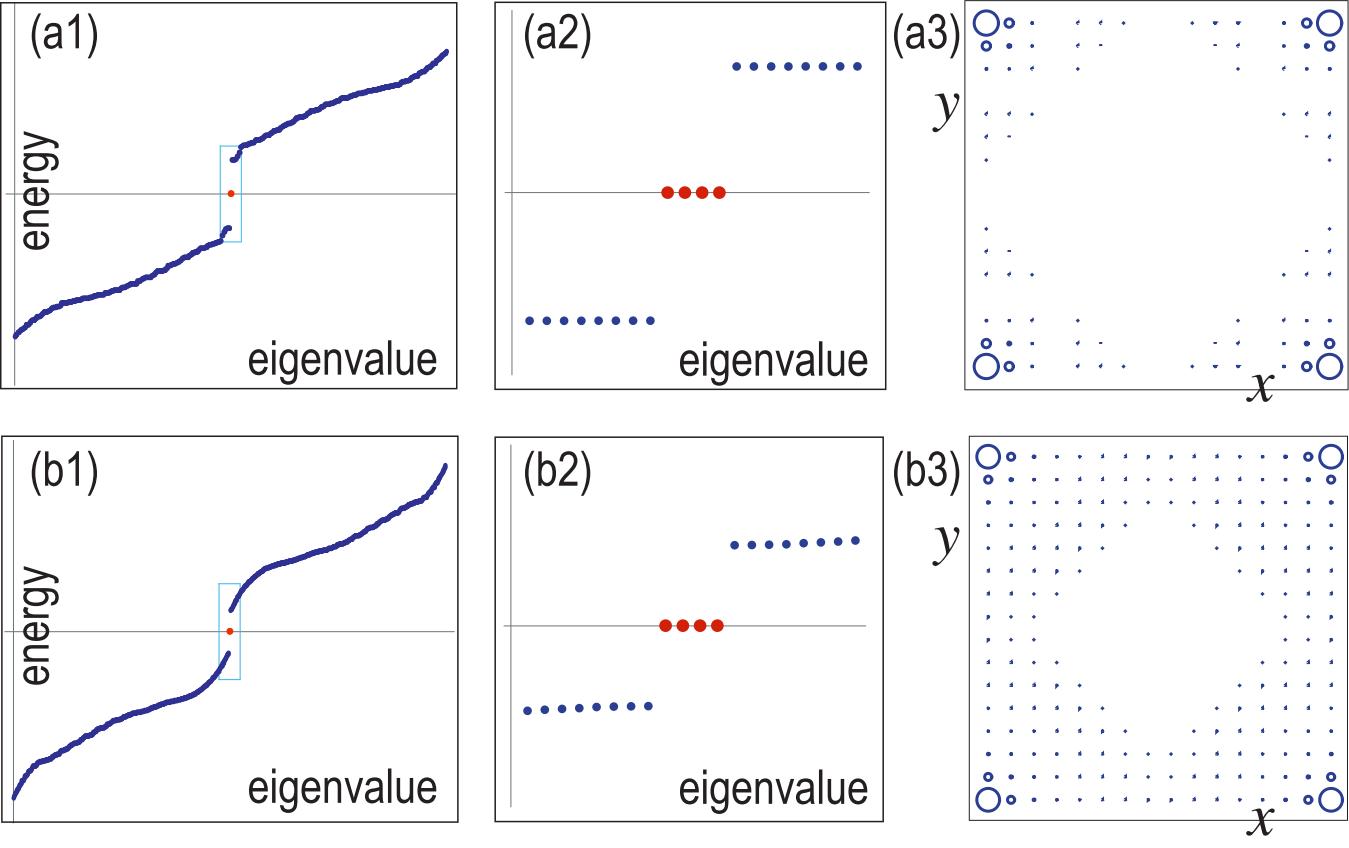}}
\caption{2D SOTI. Eigenvalues of the square (a) without magnetization and
(b) with magnetization ($B=t/2$). (a2) and (b2) Enlarged figure of the
eigenvalues corresponding to (a1) and (b1). Four zero-energy states
indicated in red are observed. (a3) and (b3) Corresponding local charge
distributions. Charge distributions are well localized, where localization
is slightly weakened in the presence of magnetization. We have set $m=%
\protect\lambda =\Delta =t$.}
\label{FigChiralDisk}
\end{figure}

\textit{Corner states:} We calculate the eigenvalues of the Hamiltonian for
a square nanodisk, which preserves the four-fold rotational symmetry. We
show the eigenvalues in Fig.\ref{FigChiralDisk}(a1) and (a2) in the absence
of the Zeeman term. There are four zero-energy states. These zero-energy
states remains as they are even when the Zeeman term is introduced as in Fig.%
\ref{FigChiralDisk}(b1) and (b2). This is a 2D magnetic SOTIs. We show the
charge distribution in the absence and presence of the Zeeman term in Fig.%
\ref{FigChiralDisk}(a3) and (b3), respectively. The wave functions are
localized at the four corners.

The origin of the gap opening in nanoribbon geometry and persistence of the
corner states in the presence of the $H_\Delta$ term is naturally understood
by treating the $H_\Delta$ as a perturbation\cite{Schin}. The edge states at
the zero energy is spatially uniform. The expectation value of the $H_\Delta$
term is $\Delta$ along the $x$ direction while it is $-\Delta$ along the $y$
direction. Thus, the edge states are gapped for a nanoribbon geometry. On
the other hand, the expectation vale of the $H_\Delta$ is exactly cancelled
at the corner. As a result, the corner states are robust in the presence of $%
H_\Delta$ term.

\begin{figure}[t]
\centerline{\includegraphics[width=0.5\textwidth]{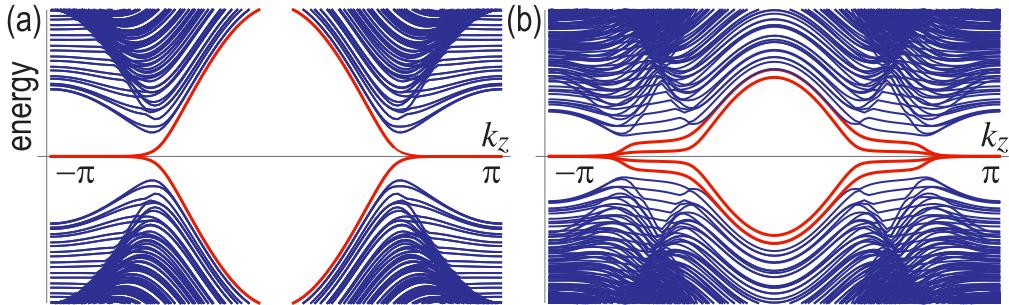}}
\caption{3D second-order topological semimetal. Band structure of hinge-arc
states (a) without the Zeeman term and (b) with the Zeeman term ($B=t/2$).
The hinge states survive even in the presence of the Zeeman term. The
horizontal axis is the momentum $k_{z}$. We have set $m/t=2$, $\protect%
\lambda =t$ and $\Delta =t/4$.}
\label{FigSM}
\end{figure}

\textbf{3D second-order topological semimetals:} By setting $%
m=m_{0}+t_{z}\cos k_{z}$ in the 2D magnetic SOTI Hamiltonian, we can
construct a model for 3D magnetic second-order topological semimetals in the
similar way to the previous works\cite{Lin,EzawaKagome}. The properties are
derived by the sliced Hamiltonian $H(k_{z})$ along the $k_{z}$ axis, which
gives a 2D magnetic SOTI model with various mass term $m$. The bulk band gap
closes at the points $k_{z}=\arccos \frac{(m-m_{0})}{t_{z}}$. The surface
states is gapped except for the two bulk gap closing points. On the other
hand, there emerge hinge-arc states connecting the two gap closing points,
which are shown in Fig.\ref{FigSM}(a). This hinge-arc states are robust even
in the presence of the Zeeman term as shown in Fig.\ref{FigSM}(b).

The author is very much grateful to N. Nagaosa for helpful discussions on
the subject. This work is supported by the Grants-in-Aid for Scientific
Research from MEXT KAKENHI (Grant Nos.JP17K05490 and JP15H05854). This work
is also supported by CREST, JST (JPMJCR16F1).

\textit{Note added:} After the completion of this work, I have become aware
of a related work\cite{Bis}, where a second-order topological insulator is
experimentally reported in Bismuth with a theoretical analysis.

\end{document}